\documentclass[10pt,preprint2]{aastex}
\def\etal{{\it et~al.}}
\def\cicam{CI\,Cam}

\def\deg{^{\circ}}

\begin{document}

\title{\cicam: A Shell-shocked X-ray Nova}
\author{Amy J. Mioduszewski\altaffilmark{1,2}}
\email{amiodusz@nrao.edu}
\author{Michael P. Rupen\altaffilmark{2}}\email{mrupen@nrao.edu}
\altaffiltext{1}{JIVE, Postbus 2, 7990 AA, Dwingeloo, the Netherlands}
\altaffiltext{2}{NRAO, P.O. Box 0, Socorro, NM 87801, USA}

\begin{abstract}
We present radio imaging observations of the 1998 outburst of the
peculiar emission line star \cicam, taken $\sim$1, 4, 75, 82, 93, 163, and
306 days after the beginning of the 31.64~March~1998 X-ray flare.
The first two epochs show a resolved but compact (no larger than 12
milliarcseconds) radio source which becomes optically thin at frequencies
higher than 5\,GHz.  The spectrum and brightness temperatures are consistent
with synchrotron self-absorption, although free-free absorption may also
play a role.  The later images show a large
(120-350\,milliarcseconds) oval-shaped or double ring remnant.  The radio 
spectrum
combined with the high brightness temperature indicates that the emission
is synchrotron, while the morphology suggests that this is powered by
a decelerating shock moving through dense circumstellar material produced by
a strong stellar wind.
The radio images of \cicam\ are equally well fit by an 
expanding ellipsoid or two expanding rings; the former gives 
$\Theta \approx 4.2 (t-50904.1)^{0.77}$, with $\Theta$ the major axis in mas
and $t$ the Modified Julian Date (MJD).  The corresponding expansion speed
in the plane of the sky was $\sim12,000\rm\,km/s$ over the first few days
(for an assumed distance of 5\,kpc), slowing by a factor $\sim3$ by the
time of the last observation almost a year later.
The radio emission from all other X-ray binary transients has either been
unresolved, or taken the form of highly collimated relativistic jets. 
We suggest that \cicam\ represents a rare case where these jets were
smothered early on by the unusually dense circumstellar medium.  In this
model \cicam\ is the analogue to extragalactic supernovae
formed by the collapsar mechanism, while the more usual X-ray binaries with
relativistic jets are analogous to the jets which escape those supernovae to
form a subset of $\gamma$-ray bursts.
\end{abstract}
 
\keywords{binaries: close $-$ stars: individual: \cicam\ $-$ radio continuum:
  stars $-$ stars: flare $-$  X-rays: binaries $-$  X-rays: bursts}
 
\section{Introduction}
 
\begin{figure}[!t]
\plotone{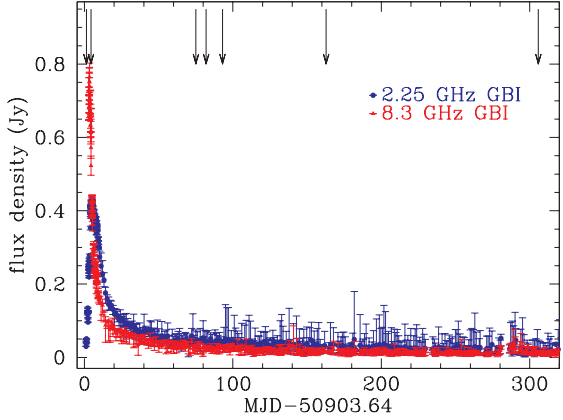}
\caption{The radio light curve from the Green Bank Interferometer (GBI),
  as produced by R.M.\ Hjellming (priv.\ comm.).
  The green triangles and red circles represent 8.3\,GHz and 2.25\,GHz
  data respectively.  After day 10 the data are averaged over 1 day and the
  error bars are the minimum and maximum for that day.  The arrows show
  the epochs of our 7 imaging observations.
}
\label{lc}
\end{figure}

XTE J0421+560 was discovered by the All-Sky Monitor (ASM) on the Rossi
X-ray Timing Explorer (RXTE) on 31.64 March 1998 \citep{Smith98}
as the first bright X-ray transient in several years, reaching almost
2 Crabs within a day of the first detection.  Corresponding radio emission
was detected soon after (1.9 April; \citealt{Hje98a}).
Optical observations \citep{WS98} identified
the radio and X-ray transient as the star \cicam , based on
positional coincidence with the radio emission, and 
a concurrent optical outburst.  It was also detected by the Burst and
Transient Source Experiment in the low--energy Gamma--rays, reaching a flux of 
1.6\,mCrab in the 20--40 keV band \citep{Har04}.

\cicam\ is a peculiar bright emission-line star with
unusually strong Fe~II lines \citep{Downes84}, indicating a very hot star or 
stellar shell.  It also has
a near-IR excess \citep{Dong91, Allen76}.  \cicam\ has a rich and unusual 
optical spectrum and has
been identified as a symbiotic star \citep{Chk70},
a Herbig~Be star \citep{The94}, a B[e] star \citep{Bel99},
and a supergiant B[e] (sgB[e]) star \citep{Rob02, Hynes02}.  
The recent consensus seems to be that it is either a B[e] or sgB[e] star.
If it is an sgB[e] star it would be the only one identified in our Galaxy, all
others having been found in the Magellanic Clouds.
Its distance has been estimated to be anything
from 1\,kpc \citep{Chk70} to 8\,kpc (e.g., \citealt{Rob02}) based on
luminosity, spectrum, interstellar absorption, and kinematics.
Here we take 5\,kpc as a reasonable compromise.  \cicam\ was not previously
a known radio source, with two 10.7\,GHz observations in 1973 giving
upper limits of 5\,mJy \citep{Alt76, Woo77}.

  \cicam's X-ray/radio/optical flare puts it into another class, that
of X-ray novae.  If \cicam\ is at a distance greater than 2\,kpc then its X-ray
luminosity indicates that there is probably a compact object (black hole or
neutron star) in the system (e.g., \citealt{Bel99}).
\citet{Orl00} suggests that the outburst was the result of
thermonuclear runaway on the surface of a white dwarf, which would require
a closer distance.
Even with all the different stellar identifications for the optical source,
it is clear that \cicam\ must be a high mass X-ray binary (HMXB).  However,
\cicam's X-ray behavior is unique among X-ray binaries.  For instance,
unlike other HMXBs, \cicam\ does not have persistent highly variable
X-ray emission, and has a soft X-ray excess in quiescence
\citep{Rob02}.  The persistent soft X-ray emission
can be attributed to the sgB[e] star and the heavily absorbed harder emission
may be an obscured compact object, but this is not certain, since this type
of emission
is extremely unusual for X-ray transients \citep{Rob02}.  The flare itself
peaked first in the X-ray, then in the optical, and finally at radio
wavelengths \citep{Fro98}, with several short, soft X-ray flares within ten
days of the outburst \citep{Fro98, Ueda98}, and rapid changes in the
X-ray absorption column over the first few days \citep{Bel99}, as well as
years afterwards \citep{Par00}.  The radio behavior is fairly typical of
X-ray transients, with an optically thick rise followed by an optically thin
decay \citep{Clark00}, but with no sign of further flaring activity.
An initial report of a relativistic radio jet \citep{Hje98b} has since been
shown to be an artifact due to a bad calibrator \citep{Rupen03}.

In this paper we present a radio imaging study of \cicam\  from a few days to
roughly a year after the X-ray outburst, using
the National Radio Astronomy Observatory's (NRAO) Very Long Baseline Array
(VLBA) and the United Kingdom national facility Multi-Element Radio Linked
Interferometer Network (MERLIN).

\section{Observations}

\begin{deluxetable}{cccccccc}
\tablewidth{0pt}
\tablecolumns{8}
\tabletypesize{\tiny}
\tablecaption{Observations}
\tablehead{
  \multicolumn{2}{c}{Epoch} &
  \colhead{days since detection} &
  \colhead{} &
  \colhead{freq.} &
  \colhead{time on \cicam} &
  \colhead{int. flux} &
  \colhead{rms}\\
 MJD\tablenotemark{a}  & date  & of X-ray flare\tablenotemark{b}
   & antennas & (GHz) & (hours) & 
density (mJy) & (mJy/beam)} 
\startdata
50904.92& 1~Apr.~98 & 1.28 & VLBA
 & 1.667 & 2.8 & 3.6\tablenotemark{c} & 0.23\tablenotemark{c} \\
50908.04& 5~Apr.~98 & 4.4 & BR FD HN LA MK NL SC Y1 & 1.667 
& 2.1 & 221.7 & 0.28 \\
50908.04& 5~Apr.~98 & 4.4 & BR FD HN LA MK NL SC Y1 & 4.987 
& 2.0 & 524.6 & 0.59 \\
50908.04& 5~Apr.~98 & 4.4 & BR FD HN LA MK NL SC Y1 & 15.37 
& 1.9 & 190.6 & 0.53 \\
50978.74& 14~June~98 & 75.1 & VLBA +Y1 & 1.667 & 1.7 & 41.5 & 0.085 \\
50985.61& 21~June~98 & 82.0 & VLBA +Y1 & 1.667 & 1.7 & 39.3 & 0.086  \\
\multicolumn{2}{c}{June~98 combined\tablenotemark{d}}& $\sim$79 &  VLBA +Y1 &
1.667 & 3.4 &40.3 & 0.075\\
50996.65& 2~July~98 & 93.0 & VLBA +Y1 & 1.667 & 2.1 & 36.1 & 0.084 \\
51066.45& 10~Sep.~98 & 162.8 & VLBA +Y1 & 1.667 & 6.8 & 20.0 & 0.036 \\
51209.44& 31~Jan.~99 & 305.8  & MERLIN & 4.994 & 10.0 & 4.3 & 0.15 \\
\enddata
\tablenotetext{a}{Modified Julian Date (Julian Date - 2400000.5) at
  mid-point of observations}
\tablenotetext{b}{The X-ray flare was first detected on 31.64 March 1998 
  (MJD 50903.64), and peaked on 1.04 April 1998 (MJD 50904.04).}
\tablenotetext{c}{Flux densities for 1~April~98 data after correction
  for the primary beam, since the VLBA was pointed 11.4 arcminutes from the
  correct position.}
\tablenotetext{d}{14~June~98 $+$ 21~June~98 data}
\tablecomments{VLBA antennas: BR-Brewster, FD-Fort Davis, HN-Hancock, 
KP-Kitt Peak, LA-Los Alamos, MK-Mauna Kea, NL-North Liberty, OV-Owens Valley,
PT-Pie Town and SC-Saint Croix; Y1 - one VLA antenna; MERLIN antennas: Mark 2 
(at Jodrell Bank), Cambridge, Defford, Knocking, Darnhall, and Tabley.}
\end{deluxetable}

\cicam\ was observed on 1~April, 5~April, 14~June, 21~June, 2~July
and 10~September 1998 with the full 10-element VLBA with the addition of one
antenna of the Very Large Array (VLA),
except for the 5~April~1998 observation, which was missing the
Pie Town, Kitt Peak, and Owens Valley VLBA antennas.  All of these
observations were carried out using left and right circular feeds
with a bandwidth of 32\,MHz in each polarization
centered on 1.7\,GHz, except for 5~April~1998,
which used three bands centered on 1.7, 5.0 and 15.4\,GHz.  MERLIN, a 6 element
interferometer in Great Britain, observed on
31 January 1999 at 5.0\,GHz, with a bandwidth of 16\,MHz in each
polarization, again using circular feeds.  Table~1 summarizes
these observations.  The VLBA observations were correlated at the VLBA 
correlator in Socorro, NM.
Figure~\ref{lc} shows the epochs of the VLBA observations in the
context of the radio light-curves, taken from the Green Bank Interferometer
(GBI) (Rupen et al., in prep.; see also \citealt{Clark00}).
The radio spectrum evolves smoothly from optically thick ($\alpha\sim 0.85$, 
$S_\nu\propto\nu^\alpha$) two days after the X-ray peak, to optically thin 
($\alpha\sim -0.4$) synchrotron radiation, 7~days after the X-ray peak.
\cicam 's radio emission is also, for X-ray novae, unusually long-lived,
 allowing 
it to be observed many months after the X-ray flare.  The lack of secondary 
radio flares, in contrast to most other X-ray novae, allows a direct link
between the radio remnant
and the original impulsive event.  \cicam\ is in fact still detected (2004) 
at the 
$\sim 1$\,mJy level with the VLA (Rupen \etal, in prep).

All the VLBA observations were phase referenced to the strong VLBI sources
J0419+5722 and J0422+5324, while the MERLIN experiment was
only referenced to J0419+5722.  Both of these sources are less 
than $1\fdg5$ away
from \cicam\ and are excellent calibrators.  For the VLBA observations
we used a cycle time of 1 minute on J0419+5722, 2.5 minutes on \cicam,
then 1 minute on J0422+5324.  The observations were reduced in the
standard manner using the NRAO's
Astronomical Image Processing System \citep{Gre03}.  The second epoch was phase
and amplitude self-calibrated, but all other epochs were not.  During the
epoch the total flux smoothly varied by 10\% in the 1.7 and 5.0\,GHz and
20\% in the 15.4\,GHz data.  The variations were flattened by the amplitude
self-calibration and did not affect the structures in the images.
The two June epochs were concatenated and imaged together.
\cicam\ evolves slowly enough that this is valid, and it improved the
image considerably.  On 1~April the VLBA observed the
estimated X-ray position, which was off by 11.4\,arcminutes. 
\cicam\ was thus at the $\sim60\%$ power point of the VLBA
antennas at 1.7\,GHz, so all flux densities for that date have been
multiplied by 1.7.
Since these observations were phase referenced the relative positions between
the high frequency the maps are accurate,
although the relative positions of the 1.7\,GHz maps may differ by up to
a few milliarcseconds (mas) due to the ionosphere. 
Also note that in these images
1\,mas$=5\,d_{5kpc}$\,AU ($d_{5kpc}=d/5$\,kpc, with $d$ the distance to the
source).  A careful polarization calibration was not done, however we can
say that \cicam\ is not polarized by more than a few percent in any epoch, 
probably less than that for the second epoch.

\begin{figure}[!t]
\plotone{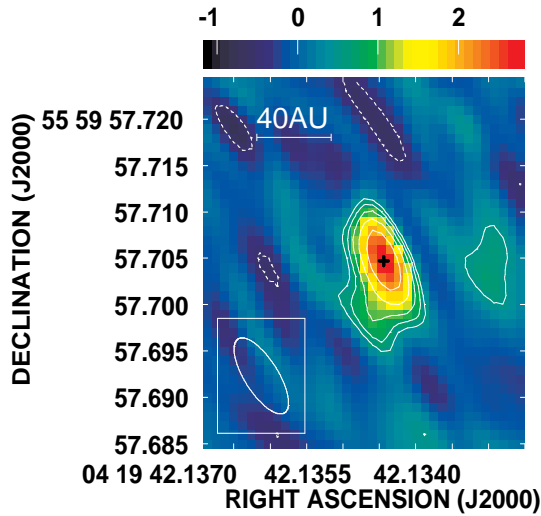}
\caption{VLBA 1.7\,GHz image of \cicam\ on 1~April~1998, less than a day 
  after the X-ray flare.  
  Peak flux density 2.77\,mJy/beam ($3.3\times10^7\rm\,K$);
  beam size $9.61 \times 3.80$\,mas at $34\fdg5$;
  rms noise 0.23\,mJy/beam;
  contours at $\pm2^{n/2}\times0.5\rm\,mJy/beam$, $n=$0,1,\dots,4; 
  conversion to brightness temperature: $12\times10^6\rm\,K/mJy/beam$.
}
\label{april1}
\end{figure}

\section{Images}
\subsection{April 1998}

Figures~\ref{april1} and \ref{april5} show the images from the first 
and second epochs.  
On 1~April~1998, during the radio rise, the source was very faint, and
still optically thick at 1.7\,GHz (see Figure~\ref{lc}).
The map shows a peak of 2\,mJy/beam; the
faint extension to the southeast (Figure~\ref{april1}) is probably a noise
spike. \cicam\ may be slightly resolved ($4\pm2\rm\,mas$) along a
position angle of $\sim 20^\circ$, but this is very close to the position angle
of the beam ($\sim 34\fdg5$), and so may not be real.
By 5~April the turnover
frequency between optically thick and optically thin emission was
about 5\,GHz, and the
images (Figure~\ref{april5}) show a much more
complicated structure.  At 1.7\,GHz, the source is resolved but very smooth,
extending $\sim$12\,mas in the north-south direction.  At optically 
thin frequencies, 5 and 15\,GHz, \cicam\ breaks up into several condensations
with a ``kidney-shaped'' structure to the south.

\subsubsection{Spectral Indices}

\begin{figure*}[t]
\epsscale{2.0}
\plotone{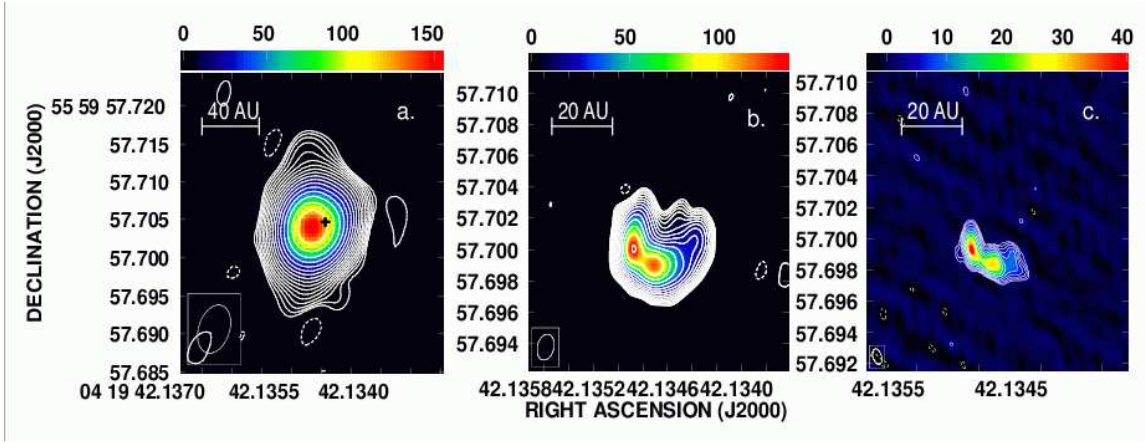}
\caption{VLBA images of \cicam\ observed on 5 April 1998.
a)~1.7\,GHz image: peak flux density 154\,mJy/beam ($2.4\times10^9\rm\,K$);
  beam size $6.58 \times 4.28$\,mas at $34\fdg5$;
  rms noise 0.28\,mJy/beam;
  contours at $\pm2^{n/2}\times0.45\rm\,mJy/beam$, $n=$0,1,\dots,14;
  conversion to brightness temperature: $16\times10^6\rm\,K/mJy/beam$.
b)~5.0\,GHz image: peak flux density 131\,mJy/beam ($3.4\times10^9\rm\,K$);
  beam size $1.70 \times 1.12$ mas at $-14\fdg3$;
  rms noise 0.59\,mJy/beam;
  contours at $\pm2^{n/2}\times2\rm\,mJy/beam$, $n=$0,1,\dots,12;
  conversion to brightness temperature: $26\times10^6\rm\,K/mJy/beam$.
c)~15.3\,GHz image: peak flux density 40\,mJy/beam ($0.35\times10^9\rm\,K$);
  beam size $1.02 \times 0.58$ mas at $18\fdg8$;
  rms noise 0.53\,mJy/beam;
  contours at $\pm2^{n/2}\times2\rm,\,mJy/beam$, $n=$0,1,\dots,7;
  conversion to brightness temperature: $8.7\times10^6\rm\,K/mJy/beam$.
}
\label{april5}
\epsscale{1.0}
\end{figure*}

  There are significant differences between the images made at different
frequencies on 5~April 1998: the 1.7\,GHz image extends further north than
the 5 and 15\,GHz images would lead one to expect, while the western arm of the
structure seen at 5\,GHz is not present in the 15\,GHz image.  Some of these
differences could easily be instrumental. Interferometers act as spatial
filters, effectively measuring different components in the Fourier transform
of the image. The Fourier component seen by a single baseline of length $B$
corresponds to a spatial frequency of $B/\lambda$.  Thus long baselines
measure rapid fluctuations of the sky brightness, while short baselines
respond to more gradual changes.  
The same physical interferometer, used at two
different wavelengths, may respond quite differently to the identical
source structure.  High-resolution interferometric images for instance may not
recover large-scale structures which contain most of the source flux density: 
one sees the crests of the waves, but not the ocean on which they rest.
For more information see \citet{SynImgII} or \citet{TMS01}.

\begin{figure}[!hbt]
\plotone{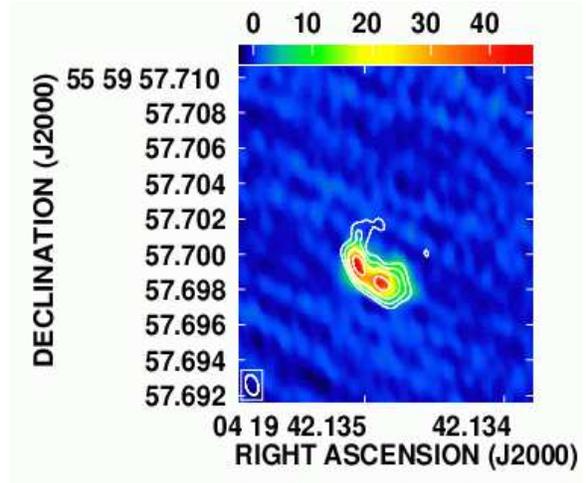}
\caption{ Contours: the CLEAN model of the 5\,GHz data,
  put through the ($u,v$)
  coverage of the 15\,GHz observation (peak flux
  density 79.0\,mJy/beam; beam size $1.18\times0.73\rm\,mas$ at 19\fdg6;
  contours at $\pm2^{n/2}\times7\rm\,mJy/beam$,
  $n=$0,2,4,6).  Colors: the observed 15\,GHz image (peak flux density
  47.2\,mJy/beam).
}
\label{spec62}
\end{figure}

  For \cicam\ on 5~April~1998, the 5\,GHz VLBA image
recovers $\sim94\%$ of the integrated flux density as measured by the VLA;
while at 15\,GHz the VLBA recovers only $\sim47\%$ of that integrated
flux density.
This strongly suggests that at least part of the 5\,GHz structure which is
missing from the 15\,GHz image has been removed by the more extensive
spatial filtering operating at 15\,GHz.  To test this idea we put the
CLEAN component model\footnote{CLEAN is a deconvolution algorithm
  which models the emission as a set of point sources; see, e.g.,
    \citet{Corn99}.}
of the 5\,GHz data through the $(u,v)$ (Fourier) coverage
at 15\,GHz, to simulate what the 5\,GHz source would look like
if observed at 15\,GHz.  This produces an image with a morphology almost
identical to that actually observed at 15\,GHz (see Figure~\ref{spec62}),
and with a total flux which is 50\% less than the same area in the 5\,GHz
image (i.e., not including the western arm).  In other words, even if its
intrinsic structure were identical to that seen in the 5\,GHz image,
the 15\,GHz
image would not show the western arm because of the spatial filtering of the
15\,GHz observation.  In addition, the observed 15\,GHz flux density would
be low by about 50\%, in agreement with the actual observations.   The only
notable distinction which remains is a $\sim14$\,mJy spur to the
northeast of the main emission in the 5\,GHz image, which is not seen at
15\,GHz.  The spectral index of this spur must therefore be somewhat steeper
than the $\alpha=-0.38\pm0.04$ ($S_\nu\propto\nu^\alpha$)
which characterizes the rest of the kidney-shaped structure.  For
comparison, the overall 5/15\,GHz spectral index on this date, as measured
with the VLA (Rupen et al., in prep.), is $-0.36$.

\begin{figure*}
\epsscale{2.0}
\plotone{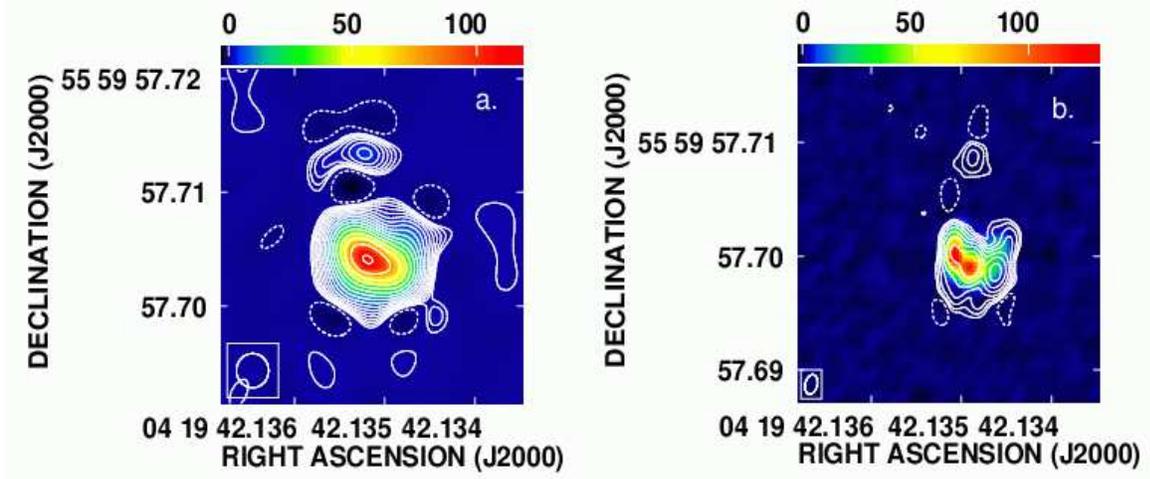}
\caption{
a)~The 1.7\,GHz image, restored with a $3\times 3$\,mas
  beam: peak flux density 1.18\,mJy/beam;
  contours same as in Figure~\ref{april5}a.
b)~Contours: the CLEAN model of the 1.7\,GHz data,
  put through the ($u,v$) coverage of the 5\,GHz observation
  (peak flux density 78.0\,mJy/beam; 
  beam size $2.03\times1.21\rm\,mas$ at $-10\fdg3$; contours at
  $\pm2^{n/2}\times0.5\rm\,mJy/beam$, $n=$0,2,\dots,12).
  Colors: the observed 5\,GHz image (peak
  flux density 131.0\,mJy/beam).
}
\label{specind}
\epsscale{1.0}
\end{figure*}

  \cicam\ is significantly more extended at 1.7\,GHz than at the higher,
optically-thin frequencies on 5~April~1998.  This is evident in
Figure~\ref{april5}, but is shown more clearly in the super-resolved
1.7\,GHz image in Figure~\ref{specind}a, where the CLEAN component model is
restored with a $3\times3$\,mas beam.  Following the same procedure as
above, Figure~\ref{specind}b shows the 1.7\,GHz model put though the 5\,GHz
$(u,v)$-coverage.  While the southern, kidney-shaped emission is fairly
similar, the northern component is seen only at 1.7\,GHz, implying a very
different spectral index.  Quantitatively, the spectral index of the
southern, kidney shaped component between 1.7 and 5\,GHz is
$\alpha=+0.78\pm0.03$, while that of the northern
component must
be steeper than $-1.0$.    This is most readily interpreted as an opacity
gradient across the source, with the northern component being much more
optically thin. 

\subsection{June 1998--January 1999}

\begin{figure*}
\epsscale{1.5}
\plotone{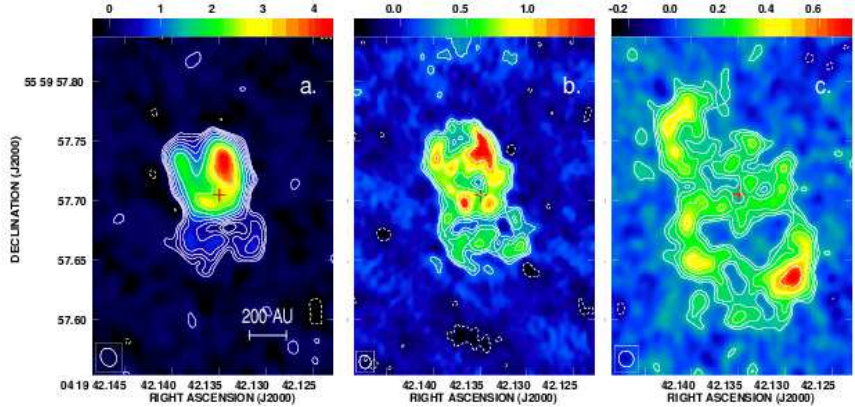}
\caption{1.7\,GHz VLBA images of \cicam\ observed months after the X-ray
flare.
  The red crosses mark the position of the peak of the 5~April~1998 1.7\,GHz
  image.
a)~Combined dataset of 14 and 21 June 1998:
  peak flux density 4.37\,mJy/beam ($8.7\times10^6\rm\,K$),
  beam size $16.13 \times 13.64$\,mas at $34\fdg5$,
  rms noise 0.075\,mJy/beam;
  contours at $\pm2^{n/2}\times0.2\rm\,mJy/beam$, $n=$0,1,\dots,6;
  conversion to brightness temperature: $2.0\times10^6\rm\,K/mJy/beam$.
  Image made with a Gaussian ($u,v$) taper of 25\,$\rm M\lambda$ (at 30\%)
  and natural weighting.
b)~Image from 2~July~1998:
  peak flux density 1.45\,mJy/beam ($5.6\times10^6\rm\,K$), 
  beam size $11.36 \times 10.03$ mas at $13\fdg3$,
  rms noise 0.084\,mJy/beam;
  contours at $\pm2^{n/2}\times0.2\rm\,mJy/beam$, $n=$0,1,2,3,4;
  conversion to brightness temperature: $3.9\times10^6\rm\,K/mJy/beam$.
  Image made with a Gaussian ($u,v$) taper of 25\,$\rm M\lambda$ (at 30\%)
  and natural weighting.
c)~Image from 10~September~1998:
  peak flux density 0.734\,mJy/beam ($1.7\times10^6\rm\,K$),
  beam size $14.49 \times 13.29$ mas at $45\fdg4$,
  rms noise 0.036\,mJy/beam;
  contours at $\pm2^{n/2}\times0.1\rm\,mJy/beam$, $n=$0,1,\dots,5;
  conversion to brightness temperature: $2.3\times10^6\rm\,K/mJy/beam$.
  Image made with a Gaussian ($u,v$) taper of 15\,$\rm M\lambda$ (at 30\%)
  and natural weighting.
}
\label{later}
\epsscale{1.0}
\end{figure*}

\begin{figure}[t]
\plotone{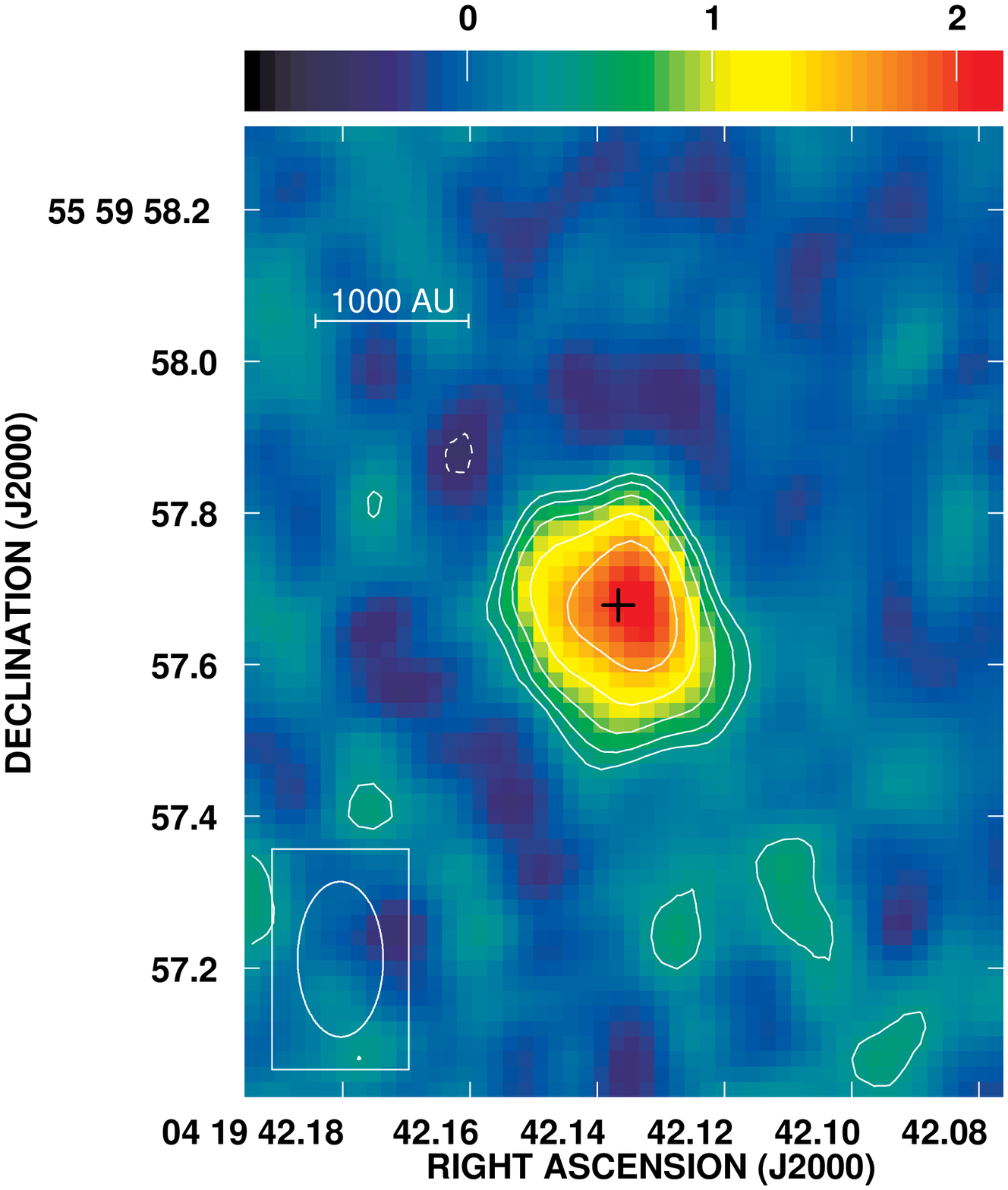}
\caption{4.99\,GHz MERLIN image of \cicam\ on 31~January~1999.
  Peak flux density 2.159\,mJy/beam ($4.6\times10^3\rm\,K$),
  beam size $204.69 \times 112.70$ mas at $0\fdg3$,
  rms noise 0.15\,mJy/beam;
  contours at $\pm2^{n/2}\times0.4\,mJy/beam$, $n=$0,1,2,3,4;
  conversion to brightness temperature: $2.1\times10^3\rm\,K/mJy/beam$.
  Image made with a Gaussian ($u,v$) taper of 2\,$\rm M\lambda$ (at 30\%)
  and natural weighting.  The black cross, as in Figures~\ref{april1},
  \ref{april5}a  and
  \ref{later}, marks the position of the peak of the 5~April~1998 1.7\,GHz
  image.  The alignment should be good to a few milliarcseconds.
  }
\label{merlin}
\end{figure}

  Images from further VLBA observations on 14-21 June, 2 July, and 10
September 1998 are displayed in  Figure~\ref{later}.  Since \cicam\ does
not evolve 
rapidly the 14 and 21 June data sets were concatenated,
significantly improving the resulting image.  Figure~\ref{merlin}
shows a MERLIN image from
31~January~1999.  The 
crosses in these images indicate the position of the peak of the 1.7\,GHz
image from 5~April~1998.   Several results are obvious from these figures.
First, the source does not bear any resemblance to a relativistic jet.
Second, \cicam\ has significantly expanded, from $\sim$12\,mas on 5 April
  1998, to at least 350\,mas in the north-south direction on
  31~January~1999, 306 days after outburst.
Third, the expansion is roughly symmetric about the center.
Fourth, there is significant emission within the source's outer rim (shell).
Finally, the shell's appearance changes significantly with time, although
  the position angle of the major axis is constant.
  In early April, the southern part is the brightest; by June-July
  the northern portion is brighter, and the structure looks almost like a
  double ring.  In September 1998 the source seems more nearly
  elliptical, and an area in the south has brightened from 1.9\,mJy in
  June to 2.7\,mJy in September, measured at the same position angle on
  the source and approximately the same area.
  Unfortunately, the MERLIN resolution is not sufficient to see any
  morphological details, but the west side is brighter than the east,
  showing that the source remains asymmetric through
  at least 31~Jan.~1999.
It is also
interesting to note that the position angle of the slightly resolved
component in the first epoch ($\sim 20^\circ$) is comparable to
that of the major axis in these later images, suggesting
that this position angle was established within a day of the outburst.

\section{Discussion}

\subsection{Emission Mechanism}

  The radio emission in \cicam\ must be primarily synchrotron, from
relativistic particles spiraling around strong magnetic fields.  This is
based on the high brightness temperatures (up to a few billion Kelvin)
seen in the VLBA images, and on the radio spectrum at high
frequencies, which follows a steeper power law than allowed for thermal
emission.  Further, any thermal plasma hot and dense enough to produce the
observed radio emission at late times would necessarily produce strong
X-ray emission as well, at levels well above what is actually observed
\citep{Par00, Orl00}.  Interestingly, the peak brightness temperature in
these resolved, optically-thick images is well below the canonical
$10^{11}-10^{12}\rm\,K$ brightness temperature limit thought to be typical of
other synchrotron sources, such as
extragalactic jets (e.g., \citealt{Kell99,Read94}) and radio supernovae
(e.g., \citealt{Wier99}).
Taking the peak of the 5~April~1998 5\,GHz image as an example,
assuming energy equipartition between relativistic electrons and the magnetic
fields, and taking the ratio of proton to electron energies as $k=100$ (as
in local cosmic rays),
  the magnetic field is $1.2d_{5\,kpc}^{-2/7}\rm\,G$,
  the total energy in relativistic particles and fields is
    $1.5\times10^{41}d_{5\,kpc}^{17/7}\rm\,erg$, and
  the synchrotron lifetime at 5\,GHz is $72d_{5\,kpc}^{3/7}\rm\,days$.
  The energy of the electrons producing synchrotron radiation at a frequency
  $\nu$ is then $30d_{\rm 5\,kpc}^{1/7}\rm (\nu/5\,GHz)^{3/7}\,MeV$.
These values should be taken as indicative rather than definitive.
Nevertheless it is clear that, as usual, the energy in relativistic particles
and fields is not an important part of the overall energy budget --- the total
X-ray emission in the flare for instance was
$\sim9.3\times10^{42}d_{5\,kpc}^2\rm\,erg$, while the total integrated
luminosity was a factor $\sim600$ higher \citep{Fro98}.  The 
rapid radio decay further shows that synchrotron losses
do not dominate the observed radio light curve, another not-unexpected
result. 

\subsection{Absorption Mechanisms}
  At early times the radio remnant is optically thick, indicating
substantial absorption.  The rapid changes in optical depth show that this
absorption is associated with the source itself, rather than some
unrelated medium.  There are two main processes which are relevant
here: synchrotron self-absorption (SSA) by the same relativistic electrons 
which
produce the emission, and free-free absorption by dense ionized gas along
the line-of-sight.  The equipartition magnetic field derived above 
would produce an SSA cutoff frequency of
    $\sim3.0d_{5\,kpc}^{2/35}\rm\,GHz$; 
below this frequency the emission should be optically thick, with a
characteristic $\nu^{2.5}$ spectrum.
This is consistent with the observed 1.7 and 5\,GHz flux densities on
5~April~1998, which imply a turnover frequency around
2.4\,GHz.  The northern component, seen only at 1.7\,GHz, could easily be
optically thin on the same date, since its much lower flux density yields
a predicted SSA turnover below 1\,GHz.  At the same time the emission
from the main, southern region must truly be as smooth as it appears in
these images: any significant clumping would lead to brightness temperatures
above $\sim\rm10^{10}\,K$, corresponding to SSA turnover frequencies
above 5\,GHz.
  
  \cicam\ is also known to have a dense stellar wind, which could lead
to substantial free-free absorption.  To produce unit opacity at 
3\,GHz requires an emission measure ($\int n_e^2\,dl$) of
  $\sim3.2\times10^7\,(T_e/10^4\rm\,K)^{1.35}\rm\,pc\,cm^{-6}$,
where $T_e$ is the electron temperature, $n_e$ is the electron number
density, and $l$ is the line-of-sight path length.  Taking an
$n_e\propto r^{-2}$ wind, assuming cosmic abundances, and taking the inner
radius of the absorbing material to be 10\,AU (corresponding roughly to
the size of the 5~April~1998 emission region), this gives a
density at 10\,AU of
  $\sim3\times10^{-18}\,(T_e/10^4\rm\,K)^{0.675}\rm\,g\,cm^{-3}$.
For comparison the modeled density for the wind in the sgB[e] star
HD\,87643 at 10 stellar radii is $10^{-16}-10^{-15}$\,g\,cm$^{-3}$ and
$10^{-12}$\,g\,cm$^{-3}$ for the hot diffuse and cold dense winds,
respectively \citep{Oud98}.  Free-free absorption could thus be quite
significant, depending on the relative orientation of the winds and the
radio-emitting region, and on the temperature and ionization of the winds.
In sum, synchrotron self-absorption must be important in the early
radio remnant, but free-free absorption may also play a role.  A detailed
analysis of the multi-frequency radio light curve may break this degeneracy.

  Even without any absorption, the radio remnant shows significant spectral
variations: the northern component seen at 1.7\,GHz on 5~April 1998 must
for instance have an intrinsic emission spectrum ($\nu^{-1}$ or steeper)
which is much steeper than the rest of the source ($\nu^{-0.4}$).
Such extreme variations have not been seen in
other shell-like sources, such as radio supernovae (Bietenholz et al., priv.\
comm.) and supernova remnants (e.g., \citealt{Wright99, Del02}).
The explanation may lie in the very
complex environment around \cicam, which could plausibly lead to 
very different densities and magnetic fields in different parts of
the circumstellar medium.

\begin{figure*}[t]
\epsscale{1.5}
\plotone{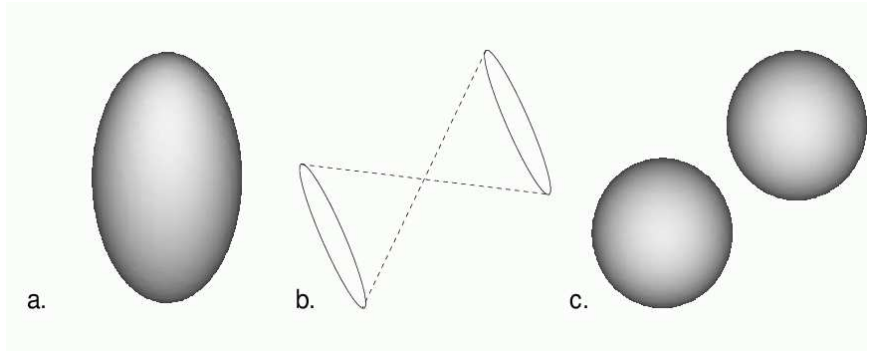}
\caption{Three possible models for the three-dimensional structure
  of \cicam.
  (a)~An ellipsoid, where patches of the surface are lit up;
  (b)~the ends of a two-sided cone (note that these must both move
    out and widen with time);
  (c)~two spheres.  
  Both the cone and the spheres, for some inclinations, could produce 
  the appearance of overlapping circles in later images (Figure~\ref{later}).
  The ellipsoid would require some patchiness in the surface, analogous to
  the brightness variations seen around the rim.  The observer is to the right.
  }
\label{scheme}
\epsscale{1.0}
\end{figure*}

\subsection{Morphology}

  The basic appearance of \cicam's radio remnant (edge-brightened, clumpy,
shell-like) strongly suggests a shock plowing into a dense circumstellar
medium.  The radio emission from all other X-ray transients has been either
unresolved, or consistent with a highly collimated one- or two-sided jet;
but there is certainly no sign of a jet in these images, and the continued
brightness of the remnant despite substantial expansion requires on-going
particle acceleration.   The relative brightness of various features changes
substantially from epoch to epoch, with some parts of the remnant even
brightening despite the overall decay; but the overall shape remains fairly
constant (see below).  The basic implication is that we are seeing the
evolving results of an explosion inside a dense, clumpy environment, much
like a radio supernova.  The changing appearance then corresponds to
changes in the local shock strength, the particle acceleration efficiency,
or the synchrotron emissivity; these in turn are presumably functions of the
local particle density and magnetic field strength and orientation.

  Even in our last image, the remnant shock is only $\sim 350$\,mas
($1750\,d_{5kpc}$\,AU) across, and is thus still running into material
dominated by \cicam's stellar wind.  The highly asymmetric and clumpy
radio morphology probably implies a similarly anisotropic and inhomogeneous
circumstellar
medium (CSM), which seems more likely than a severely asymmetric explosion or
magnetic field, and is consistent with other indications of a highly
anisotropic wind \citep{Rob02, Hynes02}.  The idea of a dense surrounding
medium is also supported by X-ray observations of \cicam\ in
quiescence, which show strong and highly variable absorption
\citep{Par00,Boi02}, and by \cicam's strong infrared
excess \citep{Dong91, Allen76}.
In this picture the changing appearance of the remnant reveals the
variations in the density of the CSM.
The initial bright kidney shaped structure to the south 
(Figure~\ref{april5}b and c) was
presumably due to a particularly dense clump on that side, which has been
over-run by the time of the June-July images (Figure~\ref{later}a and b);
while the initially fainter
emission to the north indicates a lack of dense material on
that side. 
\citet{Rob02} and \citet{Hynes02} suggest that the outflow from \cicam\
consists of a slow ($\sim30\rm\,km/s$), dense equatorial wind,
and a fast ($\sim1000\rm\,km/s$), less dense polar wind.  The shock seen
in the radio emission between 2\,July\,1998 and 10\,September\,1998 then
traces out 250/8200 years of stellar mass loss, during which time the 
density of the stellar wind changed from most dense in the north-west to 
most dense in the south-west; while the similarity between the June and
July maps shows that the wind did not vary dramatically on shorter
timescales (75/2500 years). 
Significant stellar wind variations are not unprecedented. Optical images of
ICR+10216, for instance, reveal discrete incomplete rings of dust, which
imply that the mass loss rate of the star changes on a hundred year
timescale, and is not spherically symmetric \citep{Mau99}.

  Despite these changes in brightness across the source, the overall
structure is quite similar from epoch to epoch: that is, the brightest
regions move outward at the same rate as the dim ones.  This suggests that
the radio synchrotron emission simply serves to `light up' the edges of the
shock, while remaining energetically unimportant.  By itself this is not so
surprising, as the luminosities are so low, but this further implies  that
the particle acceleration, and whatever causes it to vary around the
source, is also not dynamically important.  This is a common result for
shocks.  The radio remnant of SN\,1993J, for instance, exhibits changing
3:1 brightness asymmetries around a shell which decelerates but remains
highly circular \citep{Bie01,Bar02,Bie03}
and the supernova remnant SN\,1006 has huge
brightness asymmetries around a similarly circular rim \citep{Rog88}.

\begin{figure*}[!hbt]
\epsscale{1.5}
\plotone{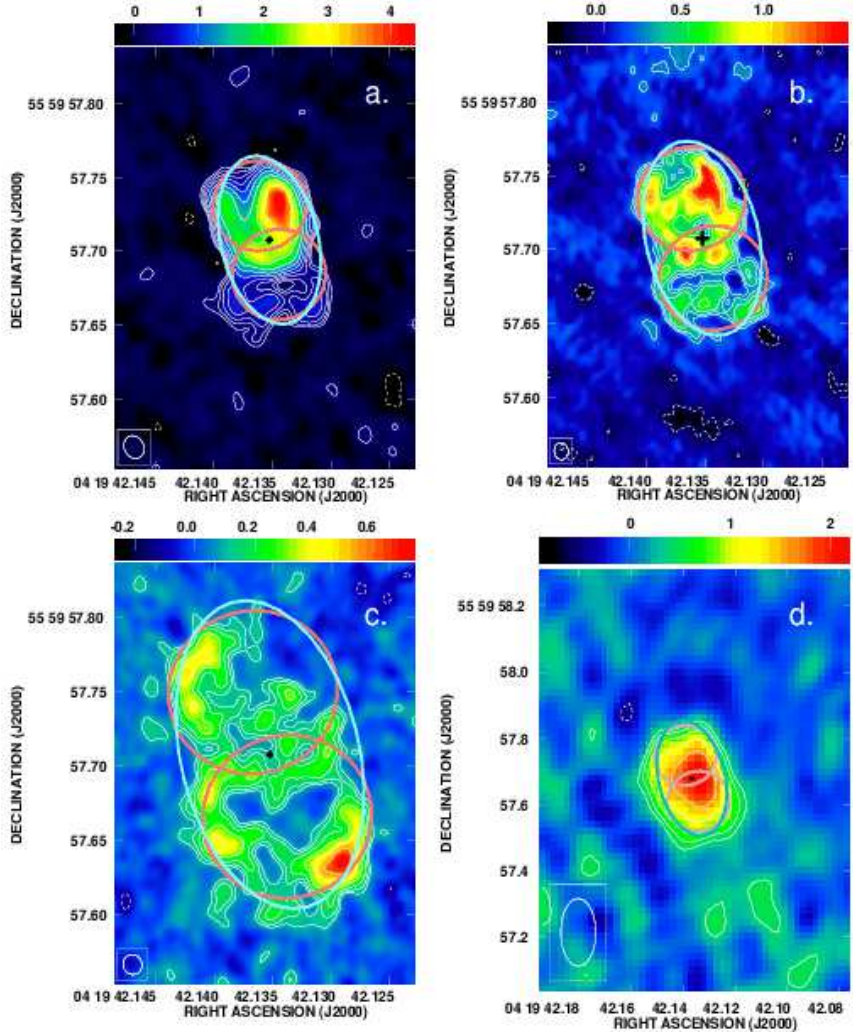}
\caption{Images from (a)~June 1998, (b)~July 1998, (c)~September 1998, 
  and (d)~January 1999, overlaid by models of a tilted ring (cyan)
  and two-sided cone (orange).  The contour levels are
  the same as in Figures~\ref{later} and \ref{merlin}.  The parameters of
  the tilted ring model are: radius $\propto t^{0.8}$, angle between normal
  and line-of-sight $53\deg\pm5\deg$,
  and position angle $15\deg$ east of north.  The parameters of the tilted
  two-sided cone model are: radius $\propto t^{0.8}$, angle between cone
  axis and line-of-sight $23\deg\pm3\deg$,
  cone opening angle $31\deg$, and position angle $7\deg$ east of north.
  }
\label{fits}
\epsscale{1.0}
\end{figure*}

\subsubsection{\cicam\ in three dimensions}

  As with most astronomical imaging, these data show a two-dimensional
projection of a three-dimensional object.  The radio remnant is clearly
elongated, so the intrinsic structure must also be asymmetric, either due to
an off-center explosion, or to an asymmetry in the surrounding medium.
One puzzle is that the source retains a similar appearance while slowing
down considerably (see below).  Several possible geometries, which could
explain the overall appearance, are shown in Figure~\ref{scheme}.

An ellipsoid could explain the outer shape of the remnant, with the brighter
regions in the middle caused by some minor asymmetry in the
particle acceleration/emission process along the line-of-sight to the
center.  However, it is difficult to retain the axis ratio of such an
ellipsoid over time. Any three-dimensional structure confined by denser
material around the `waist' would be expected to become more and more 
elongated over time, as the material in the waist should decelerate more 
rapidly than that at the poles.  Edge-brightened cone rims 
could result from a shock lighting the densest
regions of a dense wind, as in some planetary nebulae,
while a pair of spherical bubbles might arise in an explosion in a highly
asymmetric environment.
Figure~\ref{fits} shows a tilted circular ring and a two-sided tilted cone that
were fit to the data; note that the rims of the cone both move outward and
expand with time.  In both cases the expansion rate is
$\sim t^{0.8}$, as discussed in the next section.  The angle between
the normal to the titled ring and the line-of-sight is $53^\circ\pm5\deg$, while
the angle between the cone axis and the line-of-sight is $23^\circ\pm3\deg$.
These fits are illustrative rather than unique. The two-sided
cone seems to fit marginally better as it takes into account the
waist of the nebula.  The results for a pair of spherical bubbles would be
similar to those for the two-sided cone.

\subsection{Expansion}\label{sec:exp}

\begin{figure}[!hbt]
\plotone{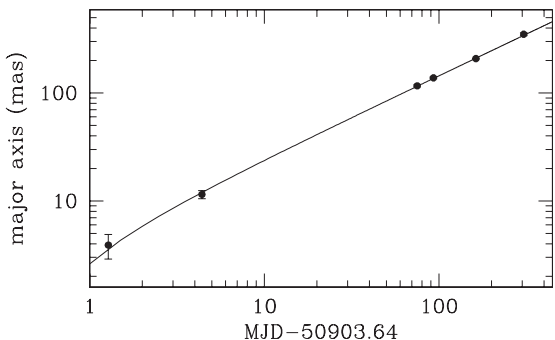}
\caption{The expansion of the major axis of \cicam\ versus
  time, where time is in days after the first X-ray detection on
  MJD\,50903.64 . 
  The major axis is estimated by fitting ellipses to the images;
  the errors are taken from
  the quality of that fit, scaled down by a factor $\sim6$ to give
  $\chi^2\sim1$ for the power-law model. 
  The five early points are taken from the  
  1.7\,GHz VLBA images shown in
  Figures~\ref{april1}, \ref{april5}a, and \ref{later}. 
  The latest point is based on the
  5\,GHz MERLIN observation shown in Figure~\ref{merlin}. The line is a
  least squares fit to the data,
    $\Theta =(4.2\pm0.9) (t-50904.09\pm0.64)^{0.77\pm0.04}$,
  with $\Theta$ in milliarcseconds and $t$ in days. 
  }
\label{exp}
\end{figure}

  The expansion rate of \cicam\ can be fit regardless of the model
used to explain the morphology, assuming that the source remains self-similar
throughout.  Figure~\ref{exp} shows the results of fitting an ellipse
to each image; the axis ratio is constant at 1.9:1, and this plot
shows the expansion of the major axis as a function of time.  The shock is
clearly decelerating, presumably because of material swept up from the
wind as the shock expands.  The average radial expansion rate of the major axis
between an assumed explosion date of 31.64 March 1998 (corresponding to the
first X-ray detection) and 5~April 1998 is $\sim1.4\rm\,mas/day$
$=11,000d_{5\,kpc}\rm\,km/s$.  This means during the 5~April~1998 observations
the source expanded by $\sim0.5\rm\,mas$.  Since the beam is bigger than this
for the 1.7 and 5\,GHz images this would have no effect.  However $\sim0.5\rm\,mas$
is comparable to the 15\,GHz beam and therefore the motion may have smeared the
image slightly in the direction of motion.  This is probably insignificant though,
since the structures in this image are $> 3\rm\,mas$.  Between the first X-ray
detection and the
MERLIN image (January 1999) the {\it average} expansion is $\sim0.6\rm\,mas/day$
$=5,000d_{5\,kpc}\rm\,km/s$.  A power-law
gives a reasonable fit to the expansion data:  
    $\Theta =(4.2\pm0.9) (t-50904.09\pm0.64)^{0.77\pm0.04}$,
where $\Theta$ is the major axis in mas, 
$t$ is the Modified Julian Date (MJD), and the error bars are $1\sigma$
assuming normally-distributed errors and forcing $\chi^2_\nu\sim1$.
The implied radial velocity is
$v=14,000 (t-50904.09)^{-0.23}\rm\,km/s$.  While only poorly determined,
the explosion date is consistent with the beginning of the X-ray flare
(on MJD 50903.64), and the fit is 
consistent with \cicam\ being marginally
resolved in the first epoch ($\Theta=3.6\rm\,mas$). 
This expansion rate ($t^{0.77}$) is much faster in than the
classic Sedov phase seen in supernovae remnants ($t^{0.4}$), but is
consistent with young supernovae (e.g., for SN\,1993J
$R \sim t^{0.92-0.78}$ [\citealt{Bar02}]).

  Given the density of \cicam's circumstellar medium, the initial
impulsive energy injection must have been quite strong to produce such a
fast shock.  One quantitative estimate comes from
the changing X-ray absorption column seen at early times \citep{Bel99}.
This
implies column densities $N$ of a few $\times10^{22}\rm\,cm^{-2}$ which decline
by more than a factor of ten over the first two days.  It seems reasonable to
associate this decline with the forcible removal of material by the shock
seen in the radio images.  By the time the X-ray absorption has decayed away 
on 2~April
\citep{Bel99} the shock was $\sim5.4\rm\,mas$
$=4\times10^{14}d_{5\,kpc}\rm\,cm$ across, according to the above
fit.  The total number of atoms swept up was $\sim N r^2$
$\sim8\times10^{51}d_{5\,kpc}^2$,
giving an average density of $\sim N/r\sim10^8d_{5\,kpc}^{-1}\rm\,cm^{-3}$, or
$\sim3\times10^{-16}d_{5\,kpc}^{-1}\rm\,g/cm^3$, assuming an average mass per
atom of 1.4 times the mass of a proton.  These values are well within the
range implied
by spectroscopy during the outburst ($<10^5$ to $>10^{10}\rm\,cm^{-3}$:
\citealt{Clark99, Orl00, Rob02, Hynes02, Mir02}), and not inconsistent with
models of the winds in sgB[e] stars (e.g., \citealt{Oud98}).  The
total mass swept up by the shock was then of order
$2\times10^{28}d_{5\,kpc}^2\rm\,g$, giving a corresponding kinetic energy
(at $13,500d_{5\,kpc}\rm\,km/s$) of
$\sim2\times10^{46}d_{5\,kpc}^4\rm\,erg$. 
This number is very uncertain, given all the underlying assumptions.
Even so it is clear that the kinetic energy in the shock could easily match
or exceed the integrated luminosity of the outburst across all observed 
wavelengths ($\sim6\times10^{45}d_{5\,kpc}^2$, \citealt{Fro98}).

  The continued slow deceleration of the shock can be used to place constraints
on the distribution of the circumstellar material via the shock jump
conditions, in a manner analogous to \citep{Chev82}'s treatment of the
expansion of radio supernovae.  
Actually what is derived is a relationship between the density structure of
the outer regions of the ejecta, and that of the CSM.  Here the `ejecta'
should be thought of as the material swept up by the time of the first
observation; while this is the stellar envelope in the case of a supernova,
for \cicam\ it is probably more appropriate to think of dense material
surrounding (accreting onto?) the compact object.
The shock between
ejecta with outer density profile $\rho_{ej}\propto R^{-n}$ and
a CSM with density profile $\rho_{CS}\propto R^{-b}$  will expand 
at a rate $R\propto t^m$, with $m=(n-3)/(n-b)$.
The observed $m=0.77$ is well within the range observed for radio supernovae
(e.g., \citealt{Bar02}), and suggests $n\sim6.3$ for an
$R^{-2}$ wind profile.  
  More generally one can argue that the observed deceleration implies that
the shock has swept up a significant amount of mass, compared to that of the
original ejecta.  The factor two drop in velocity at late times implies a
factor two to four increase in the mass over the same interval, assuming
momentum or energy conservation, respectively. 

\subsubsection{Optical Doppler shifts}

  Data taken before the April 1998 outburst showed a stable spectrum
dominated by symmetric, single-peaked hydrogen Balmer lines, with wings
extending to $\pm250\rm\,km/s$ \citep{Hynes02}.  During the outburst
both \ion{H}{1}\ and \ion{He}{1}\ lines became much broader and asymmetric,
with line wings extending up to $2500\rm\,km/s$ in H$\alpha$ and
\ion{He}{1} on 3~April~1998, and perhaps out to $\sim5000\rm\,km/s$ in the
blue wing of H$\alpha$ \citep{Hynes02}.  \citet{Rob02} also see
$\sim2500\rm\,km/s$ line wings in H and He on 9~April~1998.  While these
broad wings faded
fairly quickly \citep{Hynes02}, an ultraviolet spectrum taken
two years after the outburst showed P Cygni profiles with absorption
wings extending to $\sim-1000\rm\,km/s$ \citep{Rob02}.  It seems natural to
associate these broad lines with the outflow imaged at radio wavelengths.
The proper motions from our images correspond to a radial expansion of the
major axis at
$\sim14,000d_{5\,kpc}\,(t-50904.09)^{-0.23}\rm\,km/s$. 
The shock velocity 
was thus $\sim11,000d_{5\,kpc}\rm\,km/s$ on 3~April, and
$\sim8,600d_{5\,kpc}\rm\,km/s$ on 9~April~1998; extrapolating to two years
after the outburst gives an expected velocity of
$\sim3,100d_{5\,kpc}\rm\,km/s$.  The corresponding expansion velocities of
the minor axis of the fitted ellipse are 60\%\ of these values.

There are many uncertainties in comparing
the radio proper motions with the optical Doppler shifts.  First, the source
is clearly asymmetric, both from the radio images and from the optical line
shapes at early times, which makes the comparison of radial with transverse
velocities somewhat suspect.  Second, while the radio emission
traces the motions of the outer ejecta, the optical lines could
be sampling very different regions, with different kinematics. This is often
seen in supernovae (e.g., \citealt{Bie01}), and is clearly the case for
\cicam\ at  later times, when no substantial line wings are seen.
Given the difficulties the
rough agreement between optical and radio velocities is remarkable, and
suggests that the distance to the source is within a factor two of the nominal
5\,kpc we have adopted.  

\subsection{\cicam\ as a stellar outflow}

  There are some similarities between the radio morphology of \cicam\
and the radio outflows/nebulae observed in other
stars (e.g., AG~Pegasi [\citealt{Kenny91}] and HM~Sagittiae
[\citealt{Eyres95, Hack93, Eyres00}]),
as well as optical outflows around planetary nebulae (PNe)  
\citep{Bal02} and the high mass star Eta~Carinae \citep{Mor98}.  The
hourglass and double ring structures seen in many PNe is thought to be
caused by the density structure of the circumstellar material into which the
ejected material expands \citep{Bal02}; this is very similar to our picture
of \cicam's remnant.   However, the optical emission from PNe has a very
different physical origin from the radio synchrotron seen in \cicam, as does
the radio emission from most stellar outflows (thermal free-free radiation).
The expansion rate in \cicam\ is also much larger than the 10\,km/s
to 1000\,km/s seen in other stars (e.g., \citealt{Eyres95, Kenny91, Pal02}). 

  The overall structure of the \cicam\ radio remnant (a clumpy shell), its
initial expansion speed ($\sim 10,000\,d_{5kpc}\rm\,km/s$), and its gradual
deceleration are all very similar to what is seen in radio supernovae (e.g.,
SN\,1993J: \citealt{Bar94, Mar95, Bar02} ).  There are however two
basic differences.  First, the radio shell in \cicam\ is decidedly
asymmetric, while the few resolved radio supernovae are basically circular.
Supernova remnants on the other hand are often asymmetric, which (as in
\cicam) is most readily interpreted as the result of a shock running into a
severely anisotropic medium.  Second, while radio supernovae -- and indeed,
virtually all resolved, optically-thin radio sources -- have rather steep
radio spectra ($\alpha\sim-0.7$ to $-1.0$), \cicam\ retains a much flatter 
$\alpha\sim-0.4$ for many months.  Since the optically-thin spectral index
is a direct
consequence of the energy distribution of the relativistic electrons, this 
requires a very unusual  (but also quite stable) particle acceleration process.

\subsection{\cicam\ as an X-ray transient}
  While radio emission is often associated with X-ray transients and
binaries (e.g., \citealt{HjeHan95, Fen01}), \cicam\ is unique in showing a 
slowly-expanding, two-dimensional remnant.  In all other cases, any
resolved radio emission is dominated either by separable ejecta, or
by a collimated jet, often moving at relativistic speeds
(e.g., \citealt{Mir94, Hje95, Fom01, Mio01, Blu01, Gal04}).
This has led to the wide-spread assumption that radio emission from X-ray 
binaries always indicates the presence of a
relativistic jet, and many models now take radio emission as the best
unambiguous tracer of jet emission (e.g., \citealt{Fen03}).
\cicam\ throws this entire paradigm into doubt.  Does this have wider
implications for X-ray binaries in general? or is \cicam\ truly unique?

\subsubsection{\cicam\ as a collaspar?}
  One possibility is that \cicam\ did indeed produce a fast jet, but this
jet was quickly smothered by the dense surrounding material.   In this
model the energy in the jet was converted into a less directed outflow,
which we see as a
shock moving out into the CSM.  This is similar to the failed-jet
model for supernova explosions in collapsars (e.g., \citealt{Mac01}).
Quantitatively, one expects a relativistic jet to slow
catastrophically when it has swept up an equivalent rest mass energy
(e.g., \citealt{Rho97, Sari99}):
$$E_{jet}=\case{4}{3}\pi c^2 \left<\rho\right> r_{stop}^3$$
\noindent
where $E_{jet}$ is the energy in the jet, and $\left<\rho\right>$ is the
average density within the radius $r_{stop}$ at which the jet slows
catastrophically.  This equation is appropriate to a jet which is
stopped quickly enough that lateral expansion can rapidly isotropize the
energy.  For \cicam, we know that the remnant is far from jet-like by a
radius of 6\,mas ($30d_{5\,kpc}\rm\,AU$), and according to the arguments in
\S\ref{sec:exp} the energy in the shock (in this model derived entirely from the
original jets) is of order $\sim2\times10^{46}d_{5\,kpc}^4\rm\,erg$.
This gives a lower limit to the mean density required to stop the jets of
$\left<\rho\right>\gtrsim 10^{-19}d_{5\,kpc}\rm\,g/cm^3$, well within the
plausible range for the wind of a sgB[e] star.  It is interesting to note
that simulations of ``stalled'' jets in the context of $\gamma$-ray bursts
predict morphologies very similar to what we see in \cicam\
(see particularly the $36^\circ$-inclination images in Figures 8 and 9
of \citealt{Ayal01}).

  The question arises as to why other jets are not similarly
smothered.  Most other X-ray transients live in relatively pristine
environments, where the jet might more easily escape. In binaries with
substantial circumstellar material the accretion tends to be more continuous
(e.g., Cyg X$-$3 and SS\,433), and the resulting jets are also `on' more
often, and are thus perhaps capable of sweeping away the CSM.  This fits in 
with
\citet{Hynes02}'s picture of the \cicam\ outburst, in which a compact object
in a highly eccentric long-period ($>100\rm\,years$) orbit moves through the dense
equatorial wind of its
companion star; the brief period of super-critical accretion, when the compact
object is close to the companion, leads to the
X-ray flare and the expanding radio shock front.  A long-period orbit is required
to explain the lack of previous known flares, which would have been obvious in
the optical to historical astronomers.

\subsubsection{Shock signatures and implications for other X-ray transients}

  Given that \cicam's radio remnant is clearly dominated by a shock, it is
natural to ask whether the emission at other wavelengths could also be
associated with that shock, rather than with the accretion disk.
Extrapolation of the radio synchrotron spectrum to other wavelengths shows
that the expected synchrotron emission is far below that observed during the
outburst \citep{Bel99, Hynes02}.  Similarly the thermal emission produced by 
the
shock is likely to be quite modest, for three reasons.
First, the shock has not had enough time to become radiative, for any
  reasonable densities.  
Second, the characteristic temperature of the forward shock
  ($\sim300\rm\,keV$ for a 10,000\,km/s shock) (a)~lies outside the passband
  of most available instruments, and (b)~puts the shocked gas in a regime
  where cooling is quite slow.
Third, while the reverse shock should be slower and hence lead to lower
  and more easily observable temperatures, the resulting fluxes are
  likely to be very low; scaling from the radio/X-ray ratios in observed
  supernovae for instance (e.g., \citealt{Pool02}) the expected emission
  is orders of magnitude below the observed flux during the outburst.
The observed soft X-rays, in this as in other X-ray transients, probably arise
in and around the accretion disk rather than in the shock. 
Thus the only obvious signatures of the shock outside the radio remnant are
the optical and UV spectral lines.  Some of these are Doppler broadened by
a thousand or more km/s; others evolve very rapidly, requiring very high
densities, as might result from a shock \citep{Hynes02}.  Recent
spectra (2001-2002) still show many spectral lines at very different
strengths than their very stable pre-outburst levels \citep{Hynes02, Mir02},
which suggests some continued disturbance around the binary.
None of these is really a `smoking gun,' indicating that, for an unresolved 
X-ray transient, a shock and a freely expanding jet may not be distinguishable.

  It does not however seem likely that the radio emission from many other
unresolved X-ray binaries can be fully explained by
shocks similar to that seen here.  \cicam's radio light
curve is unusual in showing a single outburst followed by a smooth,
steep-spectrum ($S_\nu\propto\nu^{-0.4}$) decay.  Outbursts
from other sources often lead
to multiple radio events, suggesting strong on-going activity more readily
associated with the central source than with a single expanding shock wave.
There could be a series of shocks of course, but that requires re-furbishing
the surrounding medium on rather fast timescales (days to weeks).  On the 
other hand, this single shock evacuating the surrounding medium may explain why
there was only one explosive event, i.e., the shock got rid of any additional
material that could have been accreted by the compact object, and the compact
object moved away from the high density region before the stellar wind could
replenish the material.  The
quasi-steady radio emission seen in the low/hard state is also very
different from \cicam's optically-thin remnant, and requires a very small
and continuously replenished radio source.
Given these indications of very different power sources,  it is
quite surprising that \cicam\ lies nicely on the loose X-ray/radio
correlation found for X-ray binaries
\citep{Fen01, Gal03}, with a scatter typical of other X-ray transients.
It is far from obvious why direct synchrotron emission from jets should have
the same efficiency as particle acceleration in an expanding shock.

\section{Conclusions}

  We have presented evidence that the radio remnant of \cicam's 1998
outburst takes the form of a clumpy,
decelerating shell, strongly suggesting an expanding shock front, rather
than the relativistic jets seen in many other X-ray transients.  We suggest
that the reason for this distinction lies in \cicam's very dense stellar
wind, which could easily convert even quite energetic jets to a
quasi-spherical outflow.  The very broad spectral lines seen in optical
emission early on, and in the ultraviolet several months later, seem to be
the only other unambiguous signatures of this outflow.  Associating the
radio outflow with those Doppler shifts suggests that the adopted distance
of 5\,kpc is probably good to a factor of two.   Assuming the changing X-ray
absorption at early times is due to circumstellar material removed from the
line-of-sight by the radio shock, we estimate the shock energy as 
$\sim2\times10^{46}d_{5\,kpc}^4\rm\,erg$, comparable to the total integrated
luminosity of the outburst.  If the smothered jet model is correct, this
represents one of the very few measurements of a jet's total kinetic energy.
The relativistic jets seen in other X-ray binaries and transients must
either be much more powerful, or be expanding through much less dense
material.  The radio light curves of most unresolved sources suggest
continued emission from a small region, so it also seems unlikely that extended
shocks dominate the radio emission for these sources.   \cicam\ seems to be
related to other X-ray transients, as collapsar supernovae are related to
$\gamma$-ray bursts.

\acknowledgments
This paper would not have existed without Robert Hjellming, who
provided the drive to observe this source with the
VLBA even before we knew it was a strong radio source.  He also 
edited the first 40 days of the GBI data.  We are further indebted to Vivek
Dhawan, Kristy Dyer, Dale Frail, and Rob Hynes for useful discussions.
The VLA and the VLBA are facilities of
NRAO which is operated by
Associated Universities, Inc. under cooperative agreement with
the National Science Foundation.
The GBI is a facility
of the National Science Foundation operated by the National Radio Astronomy
Observatory (NRAO) in support of NASA High Energy Astrophysics programs.
A.J.M. acknowledges support from the European Commission's
TMR/LSF Programme (Contract No. ERB-FMGE-CT95-0012).
Basic research
in radio astronomy at the Naval Research Laboratory is funded
by the Office of Naval Research.  This research has made use of
the SIMBAD database, operated at CDS, Strasbourg, France, and
NASA's Astrophysics Data System.

\end{document}